# VoIP Steganography and Its Detection – A Survey

WOJCIECH MAZURCZYK, Warsaw University of Technology

Steganography is an ancient art that encompasses various techniques of information hiding, the aim of which is to embed secret information into a carrier message. Steganographic methods are usually aimed at hiding the very existence of the communication. Due to the rise in popularity of IP telephony, together with the large volume of data and variety of protocols involved, it is currently attracting the attention of the research community as a perfect carrier for steganographic purposes. This paper is a first survey of the existing VoIP steganography methods and their countermeasures.

## 1. INTRODUCTION

IP telephony or Voice over IP (VoIP) is a real-time service which enables users to make phone calls through IP data networks. It is one of the most important services of IP-based networks and is impacting the entire telecommunications landscape.

An IP telephony connection consists of two phases in which certain types of traffic are exchanged between the calling parties: signalling and conversation phases. During the first phase certain signalling protocol messages, for example SIP (Session Initiation Protocol) messages [Rosenberg et al. 2002], are exchanged between the caller and callee. These messages are intended to set up and negotiate the connection parameters between the calling parties. During the second phase two audio streams are sent bi-directionally. RTP (Real-Time Transport Protocol) [Schulzrinne et al. 2003] is most often utilised for voice data transport and thus packets that carry the voice payload are called RTP packets. The consecutive RTP packets form an RTP stream.

Steganography is an ancient art that encompasses various information hiding techniques, whose aim is to embed a secret message into a carrier (steganogram). Steganographic methods are aimed at hiding the very existence of the communication and therefore keep any third-party observers unaware of the presence of the steganographic exchange. Steganographic carriers have evolved through the ages and are related to the evolution of the methods of communication between people. Thus, it is not surprising that current telecommunication networks are a natural target for steganography and in particular, IP telephony is attracting the attention of the steganography research community. It is because of the following features that IP telephony is a perfect carrier for steganographic purposes:

— It is very popular, thus its usage will not raise suspicions, i.e., it will not be considered as an anomaly itself.
— The large volume of VoIP data. The more frequent the presence and utilisation of such carriers in networks, the better their masking capacity, as hidden communications can pass unnoticed amongst the bulk of exchanged data.
— Potentially high steganographic bandwidth that can be achieved. For example, during the conversation phase of a G.711-based call, each RTP packet carries 20 ms of voice; in this case the RTP stream rate is 50 packets per second. Thus, even by simply hiding 1 bit in every RTP packet we gain quite a high steganographic bandwidth of 50 bit/s.

This work was supported by the Polish Ministry of Science and Higher Education and Polish National Science Centre under grants: 0349/IP2/2011/71 and 2011/01/D/ST7/05054.

Author's addresses: W. Mazurczyk, Warsaw University of Technology; Institute of Telecommunications, Warsaw, Poland, 00-665, Nowowiejska 15/19, email: wmazurczyk@tele.pw.edu.pl.

— It involves the combined use of a variety of protocols. Thus, many opportunities for hiding information arise from the different layers of the TCP/IP stack. Hidden communication can be enabled by employing steganographic methods applied to the users' voice that is carried inside the RTP packets' payload, by utilising so called well-known digital media steganography, or by utilising VoIP protocols as a steganographic carrier. This makes VoIP a multi-dimensional carrier.
— It is a real-time service, which induces additional strict requirements for steganographic methods and its detection (steganalysis) ones. It also simultaneously creates new opportunities for steganography (e.g. utilisation of excessively delayed packets that are discarded by the receiver without processing, because they cannot be considered for voice reconstruction).
— The VoIP calls are dynamic and of variable length which make VoIP-based steganography even harder to detect.

Presently, steganographic methods that can be utilised in telecommunication networks are jointly described by the term network steganography or, specifically when applied to IP telephony, by the terms VoIP steganography or steganophony [Lubacz et al. 2008]. These terms pertain to the techniques of hiding information in any layer of the TCP/IP protocol stack (Fig. 1), including techniques applied to the speech codecs, or those that utilise the speech itself.

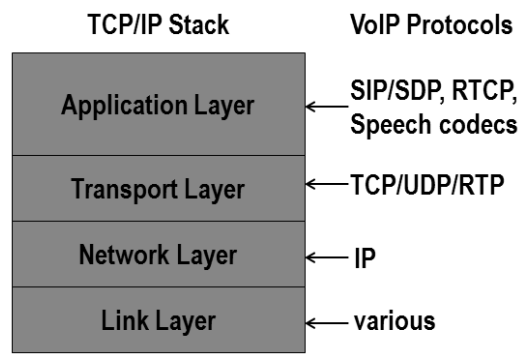

Fig. 1. VoIP stack and protocols.

In general, steganography can be treated as a double-edged sword depending on who uses it and how. However, the ethical issues related to the utilisation of information hiding techniques require consideration in a broader steganography context, which is beyond the scope of this paper. The main application of network steganography is for providing a means of conducting clandestine communication. The purposes for establishing hidden communication can be varied; possible uses can fall into the category of legal or illicit activities. The illegal aspect of steganography starts from criminal communication, through confidential data exfiltration from guarded systems, cyber weapon exchange and control, up to industrial espionage. Legitimate uses include circumvention of web censorship and surveillance, computer/network forensics, or copyright protection. Techniques can be used with VoIP to improve its resistance to packet losses and improve voice quality [Aoki 2003; 2004], extend communication bandwidth [Aoki 2007] or provide means for secure cryptographic key distribution [Huang et al. 2011a].

The expansion of TCP/IP networks opened up many possibilities for covert communication due to changes in the traditional circuit-switched networks paradigm; services/applications are created by the network users rather than by the network itself, the transport and control functions are not separated and can be influenced by the user. These possibilities are a consequence of the fact that network users can influence, and/or use the control of data flow – the communication protocols – together with the service/application functionality of terminals to establish covert communication. That is why secret messages can be hidden, not only within ordinary non-covert (overt) messages as in traditional steganography and circuit-switched networks, but also in the communication protocol's control elements, in effect by manipulation of the protocol's logic, or by combinations of the above.

A considerable number of steganographic methods have been developed so far [Zander et al. 2007; Bender et al. 1996] and they cover all layers of the TCP/IP stack. Moreover, many of the previously proposed steganographic methods can be successfully adapted to VoIP traffic and many new VoIP-specific methods have been introduced and they are covered in this survey. However, if we look at VoIP from the perspective of traffic, then in general, every VoIP steganography technique can be classified into one of the three following groups (Fig. 2) [Lubacz et al. 2008] based on what is used as the steganographic carrier:

— S1: steganographic methods that modify protocol PDU (Protocol Data Unit) – network protocol headers or payload field. Examples of such solutions include: (1) modifications of free/redundant headers' fields of IP, UDP or RTP protocols during conversation phase and (2) modification of signalling messages in e.g., SIP (using the same principle as in [Murdoch and Lewis 2005]), or (3) modification of the RTP packets' payload by modifying the user content that it carries (e.g., by employing widely used Least Significant Bits modification method [Bender et al. 1996]), or simply by replacing user data [Mazurczyk et al. 2011]. Obviously, methods that rely on modification of both the header fields and payload are also possible.
— S2: steganographic methods that modify PDUs' time relations, e.g. by modifying PDUs inter-packet delay [Wang et al. 2005], [Shah et al. 2006], by affecting the sequence order of PDUs (similarly like in [Kundur and Ahsan 2003]) or by introducing intentional PDU losses (by adopting solution from [Servetto and Vetterli 2001]).
— S3: Hybrid steganographic methods that modify both the content of PDUs and their time relations. An example of such a solution is the LACK (Lost Audio Packets Steganography) method [Mazurczyk and Szczypiorski 2008], which will be described in Section 3.

This classification quite precisely describes current information hiding possibilities in VoIP. However, it must be noted that the main focus of the research community in the last decade was dedicated to the methods from the S1 group. Solutions from the S2 group are harder to deploy practically, because they usually offer low steganographic bandwidth and require synchronisation. Moreover, they can significantly degrade the quality of the call. Methods from the S3 group are rather recent discoveries, thus to date, there have not been many papers on this subject.

It must also be noted that there are a number of information hiding techniques proposed for certain codecs, or audio files that could be also utilised for VoIP (for review of audio steganography methods interested readers are referred to the book edited by Lu [2005]). Also there are number of other existing steganographic methods e.g. protocol steganography ones that can be potentially utilized in VoIP environment. However, many of these have not been evaluated for VoIP per se, thus, they are intentionally omitted. Therefore, it must be emphasised that in this paper we focus **only** on these existing steganographic and detection methods that have been proved feasible for VoIP.

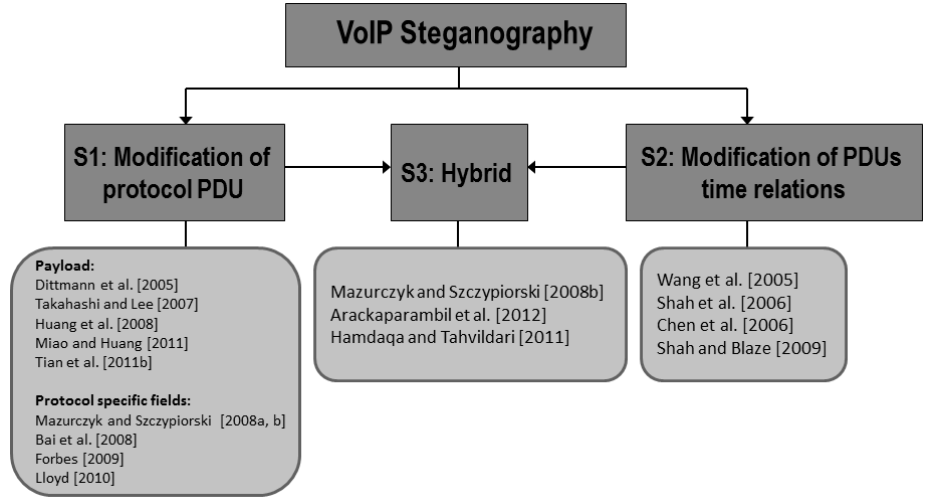

Fig. 2. VoIP steganography classification with exemplary methods

This paper is a first review of the practical, existing VoIP steganography methods and countermeasures available during the period 2003-2012. The rest of the paper is organised as follows. Section 2 introduces the fundamentals of network steganography and VoIP steganography in particular. Section 3 describes the existing steganographic techniques, whilst Section 4 presents possible countermeasures and finally, Section 5 concludes this work.

## 2. STEGANOGRAPHY: TERMINOLOGY, COMMUNICATION MODEL AND SCENARIOS

As mentioned in Section 1, the means of communication amongst people has evolved over time; from messengers, letters (e.g., where sympathetic ink was used to facilitate hidden data exchange) and telephones, to computer networks and simultaneously, so have steganographic methods evolved. Therefore, techniques of hiding information that utilise network protocols can be viewed as an evolutionary step of the hidden data carrier, rather than some new phenomenon.

The scientific community has been using many terms such as steganography [Petitcolas et al. 1999; Fridrich 2010], covert channels [Lampson 1973; DoD 1985], or information hiding [Petitcolas et al. 1999] to describe the process of concealing information in the digital environment. This stems from the fact that the terms have not been introduced at the same time and because their definitions have evolved. However, in our opinion drawing a distinction between steganography and covert channels in telecommunication networks environment, is not necessary and instead, one term – network steganography – should be used. It is our belief that in this case steganographic methods are used to create a covert (steganographic) channel. Thus, the scope of network steganography encompasses all information hiding techniques that can be applied in telecommunication networks to enable hidden data exchange.

The best example to prove this point is in fact IP telephony services. Due to its nature, information hiding is possible within the speech that is carried inside the voice packets and in the modification of the protocols that enables it. The carrier is actually the same; the VoIP service and the methods used in both cases are inseparably bound to the process of communicating through the network. That is why, when considering steganography applied to the VoIP service, terms such as; VoIP steganography or steganophony should be used.

Another argument against using two separate terms is that in the state of the art for steganography [Petitcolas et al. 1999] as well as for covert channels [Zander et al. 2007] the same model is referenced as a hidden communication model. It is the famous "prisoners' problem" which was first formulated by Simmons [1984] in 1983 (Fig. 3). In this model two prisoners; Alice and Bob, are jailed in separate cells and they are trying to prepare an escape plan. The problem is that their communication is always passed through and inspected by the Warden. If the Warden identifies any conspiracy, he will put them into solitary confinement, so Alice and Bob must find a way to exchange hidden messages for their escape plan to succeed. The solution is to use steganography. By concealing a hidden message ($M_{HID}$) in an innocent looking carrier ($M_{CAR}$), it is possible to achieve a modified carrier ($M_{STEG}$) that will raise no suspicions while traversing the communication channel. Thus, to be able to create a covert channel in which hidden data is exchanged, the utilisation of a steganographic method by the sender ($F_{STEG}$ in Fig. 3) and by the receiver ($F^{-1}_{STEG}$ in Fig. 3) is always necessary. For Alice and Bob, the communication channel is also a covert channel that was created using given a steganographic method.

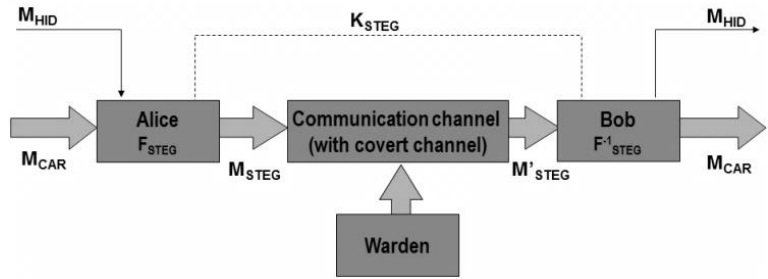

Fig. 3. Model for hidden communication.

The similarity between these two terms (steganography and covert channel) was in fact also observed by other researchers in the field (e.g., by Fridrich [2010]).

For steganographic methods, it is usually assumed that there exists a secret stego-key ($K_{STEG}$ in Fig. 3) that is a form of shared secret between the communicating parties and allows to build security of steganographic system. In network steganography, knowledge of how the information hiding technique operates can be also treated as a stego-key. Therefore, everything else related to the steganogram transmission can be known to the warden – the entity that performs detection (steganalysis). In particular it:

— is aware that Alice and Bob can be utilising hidden communication to exchange data in a covert manner.
— has a knowledge of all existing steganographic methods, but not of the one used by Alice and Bob (this, as mentioned earlier, is assumed to be their stego-key).
— is able to try to detect, and/or interrupt the hidden communication.

For VoIP steganography, four possible hidden communication scenarios may be considered, as illustrated in Fig. 4. The first scenario (marked with 1 in Fig. 4) is most common: the sender and the receiver perform VoIP conversation while simultaneously exchanging steganograms. The conversation path is the same as the hidden path. For the next three scenarios (marked 2-4 in Fig. 4) only a part of the VoIP end-to-end path is used for the hidden communication, as a result of actions undertaken by intermediate nodes; the sender and receiver are, in principle, unaware of the steganographic data exchange. The possible localisations of the warden are denoted as W1-W3 and will be discussed in detail in Section 4.

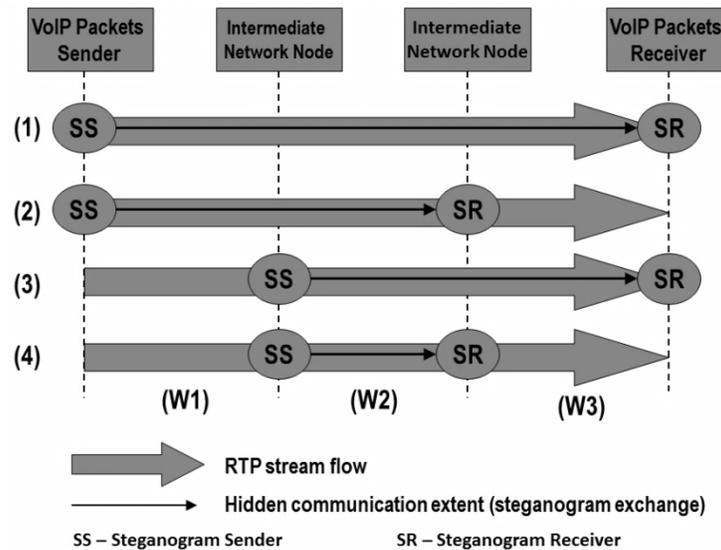

Fig. 4. Model for hidden communication (based on [Lucena et al 2004] and [Zander et al. 2007]).

Every network steganographic method can be described typically by the following set of characteristics: its *steganographic bandwidth* (also referred as capacity typically for media steganography), its *undetectability* (also referred as security in literature Fridrich [2010]), and *robustness*. The term "steganographic bandwidth" refers to the amount of secret data that can be sent per unit time when using a particular method. Undetectability is defined as the inability to detect a steganogram within a certain carrier. The most popular way to detect a steganogram is to analyse the statistical properties of the captured data and compare them with the typical values for that carrier. The last characteristic is robustness that is defined as the amount of alteration steganogram can withstand without secret data being destroyed.

Of course a good steganographic method should be as robust and hard to detect as possible while offering the highest bandwidth. However it must be noted that there is always a fundamental trade-off among these three measures necessary.

Additionally, it is useful also to measure the steganographic cost. It is a characteristic that belongs to the sphere of carrier fidelity and has direct impact on undetectability. It describes the degradation or distortion of the carrier caused by the application of the steganographic method: similarly as MSE (Mean-Square Error) or PSNR (Peak Signal-to-Noise Ratio) were utilized for digital media carriers. In the case of VoIP steganography methods, this cost can be expressed, for example, by providing a measure of the conversation quality degradation induced by applying a particular information hiding technique.

## 3. VOIP STEGANOGRAPHY

IP telephony as a hidden data carrier can be considered a fairly recent discovery. However, existing VoIP steganographic methods stem from two distinct research origins.

The first is the well-established image, audio and video file steganography (also called digital media steganography) [Bender et al. 1996], which has given rise to methods that target the digital representation of the transmitted voice as the carrier for hidden data.

The second sphere of influence is the so-called covert channels (comp. Section 2), created in different network protocols [Kundur and Ahsan 2003; Murdoch and Lewis 2005] (a good survey on covert channels can be found in [Zander et al. 2007]), also referred in literature as protocol steganography [Lucena et al. 2004]; these solutions target specific VoIP protocol fields (e.g. signalling protocol – SIP, transport protocol – RTP, or control protocol – RTCP), or their behaviours.

In the following subsections we analyse VoIP steganography methods taking into account the two research origins mentioned above. In each case chronology is retained.

### 3.1 Steganographic methods applied to voice payload

All information hiding methods described in this subsection belong generally to the group S1 from classification presented in Fig. 2 (modifying the payload of the VoIP PDU's).

#### 3.1.1. LSB-based methods applied to digital voice signal

Surprisingly, a lot of research effort is still devoted for improving LSB (Least Significant Bit) steganography. However, it must be noted that some of the techniques described below can be successfully applied also to other VoIP steganography methods to increase their undetectability, robustness or steganographic bandwidth.

The first VoIP steganographic method to utilise the digital voice signal as a hidden data carrier was proposed by Aoki [2003] in 2003. The LSB steganography was utilised to provide PLC (Packet Loss Concealment) method for G.711-based VoIP. Later, the PLC method was further improved ([Aoki 2004; 2007]), but the steganographic method and voice codec remained the same.

Dittmann et al., [2005] in 2005 presented the first VoIP steganography implementation prototype that also used the LSB method. This work was further extended and published in 2006 [Krätzer et al. 2006] by demonstrating testing which proved that a typical VoIP communication can be practically used for steganographic purposes.

Wu and Yang [2006] described the scheme of adaptive LSB, which for G.711-based speech calculates energy statistics to estimate the number of least significant bits to be utilised as a hidden data carrier in each voice sample. The results proved that this approach performs better than simple LSB and offers a higher steganographic bandwidth (about 20 Kbps) while introducing less degradation of the voice quality.

In [Druid 2007], an implementation of SteganRTP was described. This tool employs the least significant bits of the G.711 codec to carry steganograms and offers a reliable bi-directional covert communication channel that allows to exchange 1 kB/s of secret data in single direction.

Wang and Wu [2007] also suggested using the least significant bits of voice samples to carry secret communication, but in their solution the bits of the steganogram were coded using a low rate voice codec, like Speex. Their prototype implementation is characterised by a small processing delay of about 0.257 ms.

Takahashi and Lee [2007] presented a proof of concept LSB-based implementation, Voice over VoIP (Vo$^2$IP), which is able to establish a hidden communication by embedding 8kbit/s G.729-based compressed voice data into the regular, PCM (Pulse Code Modulation)-based, voice traffic.

Liu et al. [2008] found that the LSBs of each speech frame for G.729 can be replaced with secret data bits. The experimental results indicate that the method is perceptually transparent while the steganographic bandwidth is relatively high (about 200 bit/s).

Huang et al. [2008] described how to use the LSB matching method to enable covert communication for VoIP. They developed a G.711-based prototype called Stega-Talk that also offers the recovery of secret data lost due to network conditions. It is based on the Redundant Audio Data RTP payload type that is typically used to carry a DTMF (Dual-Tone Multi-Frequency) digit.

Tian et al. [2008] proposed the use of an LSB steganography-based system that employs a well-balanced and simple encryption of secret data. This system was evaluated for VoIP with G.729a speech coding using a proof of concept tool named StegTalk. The experimental results showed that the achievable steganographic bandwidth is in the range 0.8–2.6 kbit/s and has a negligible effect on speech quality. Moreover, it met the real-time requirements of the VoIP service.

A real-time steganography system for VoIP was described by Tian et al., [2009a]. The main novelty of the proposed solution is not in the steganographic method used (LSB), but in utilising M-sequence encryption techniques to eliminate the correlation amongst secret messages, to increase the resistance against statistical steganalysis. Moreover, protocol steganography (usage of free/unused fields in protocols headers) is used to provide a novel synchronisation mechanism together with an RSA-based key agreement that ensures accurate restitution of the secret messages on the receiver side. This system was experimentally evaluated for 0.8 and 2.6 kbit/s steganographic bandwidth, which obtained a 0.3 and 1 quality drop in the MOS (Mean Opinion Score) scale (which is typically used for expressing the quality of VoIP calls), respectively and the total embedding latency increased by about 4.7 ms when 1 MB of steganogram is transmitted. A similar LSB-based approach, relying on adaptive VoIP steganography was presented by the same authors in [Tian et al. 2009b].

Xu and Yang [2009] proposed an LSB-based method dedicated to voice transmission using the G.723.1 codec in 5.3 kbit/s mode. They identified five least significant bits of the LSP VQ (Line Spectrum Pair Vector Quantization) indices and used them to transmit hidden data; the method provided a steganographic bandwidth of 133.3 bit/s.

Tian et al. [2010] described a Dynamic Matrix Encoding Strategy (DMES) that dynamically chooses the size of each message group in a given set of adoptable message sizes. The purpose of DMES is to flexibly adjust the steganographic bandwidth and embedding transparency in accordance with the requirements of the users. Its main advantage is that it is also codec and cover independent.

An adaptive steganography scheme based on the smoothness of the speech block was introduced by Miao and Huang [2011]. By choosing lower embedding rates in the flat blocks and higher in sharp ones the security of the method was improved. Such an approach outperforms the classic LSB-based methods in terms of voice quality. Experimental results showed that about 7.5 kbit/s of secret data can be sent covertly with a degradation of voice quality of less than 0.5 (expressed again on the MOS scale).

Tian et al., [2011a] presented an Adaptive Partial-Matching Steganography (APMS) approach. They introduced a partial similarity value (PSV) measure to evaluate the partial matching between covert and overt messages. This allows an adaptive balance of the steganographic transparency and bandwidth. Additionally, they utilise triple M-sequences to eliminate the correlation among secret messages, guide the adaptive embedding process, and encrypt synchronisation signalling patterns.

An insightful overview of the general techniques that can be applied to VoIP steganography methods to make their detection even more difficult was introduced by Tian et al., [2011b]. Additionally, they proposed three new encoding strategies based on digital logic. All techniques were evaluated for LSB-based steganography and proved to be effective.

Another adaptive LSB-based steganography approach named AVIS (Adaptive VoIP Steganography) was proposed by Xu et al. [2011]. AVIS has two components: VAMI (Value-based Multiple Insertion), which is responsible for dynamically selecting multiple bits based on the VoIP vector value and VADDI (Voice Activity Detection Dynamic Insertion), which dynamically changes the embedding intervals to make detection harder. The approach was implemented for G.711-based VoIP and the results prove that it is less detectable than a classic LSB method whilst achieving a steganographic bandwidth of about 114 B/s, introducing acceptable delay and degrading the voice from 0.1 to 0.4 on MOS scale.

Liu et al. [2012] adopted least-significant-digits rather than LSBs to hide secret data. This approach can increase around 30% of steganographic bandwidth while introducing lower steganographic cost than classic LSB method.

### 3.1.2. Other methods

Besides LSB-based information hiding approaches other effective methods that utilize speech as a secret data carrier we proposed. These solutions are generally based on:

- Phase coding ([Takahashi and Lee 2007], [Nutzinger and Wurzer 2011]),
- QIM (Quantization Index Modulation) technique ([Xiao et al. 2008]),
- Spectrum techniques in transform domain ([Takahashi and Lee 2007], [Nutzinger et al. 2010]),
- Echo hiding technique ([Takahashi and Lee 2007]),
- Analysis by synthesis (ABS)-based scheme ([Ma et al. 2007], [Wu et al. 2009]),
- DFT (Discrete Fourier Transform) technique ([Deng et al. 2008])
- Speech codec-specific approaches ([Aoki 2008, 2009, 2010], [Geiser et al. 2008], [Nishimura 2009], [Huang et al. 2011b]),

Takahashi and Lee in [2007] besides LSB method considered the feasibility of other methods that can be utilised in VoIP steganography, like DSSS (Direct Sequence Spread Spectrum), FHSS (Frequency-Hopping Spread Spectrum), or Echo hiding. They also provided experimental results of steganographic bandwidth, which was about 20 bit/s for all three techniques, while still maintaining good voice quality robustness. Nutzinger et al. [2010] further improved DSSS by creating a hybrid steganographic algorithm that combined DSSS with frequency hopping and bit-rate variation. Authors also implemented a prototype and stated that the proposed scheme has none to negligible influence on the speech quality.

Ma et al., [2007] adopted an ABS (Analysis-By-Synthesis) algorithm [Wu et al. 2003] to enable hiding of 2.4 kbit/s MELP (Mixed-Excitation Linear Prediction) speech in G.721-based overt speech. The obtained steganographic bandwidth was estimated up to 8 kbit/s. Also ABS was utilized by Wu et al. [2009] to build an information hiding scheme based on LPCs (Linear Predictive Coefficients) as the secret data carrier by means of LPCs substitution. The experimental results for four speech codecs: G.721, GSM, G.728 and G.729 proved that the proposed approach is characterised by high steganographic bandwidth (from 800 to 3600 bit/s) while offering good undetectability, robustness and ability to perform in real-time. The results were also compared with four traditional information hiding technologies: LSB, echo hiding, phase encoding and spectrum transform and also showed to be superior.

A steganographic method based on the characteristics of PCMU (PCM μ-law) in which the 0-th speech sample can be represented by two codes due to the overlap (namely +0 and -0) was proposed by Aoki [2008]. This redundancy is then exploited to embed hidden data into the speech. This technique is lossless, thus there is no voice degradation. The obtained steganographic bandwidth of this method was estimated to be in the range 4.4–24 kbit/s depending on the extent of the background noise level. The work was later extended and published in 2009 [Aoki 2009] and studied lossless steganography techniques for G.711 PCMU and DVI-ADPCM (DVI Adaptive Differential Pulse Code Modulation) codecs. The proposed techniques also exploit the redundancy of the folded binary code employed in these codecs for embedding a steganogram into the speech without quality degradation. The improved technique offers steganographic bandwidth for G.711-based calls from 24 to 400 bit/s and from 0 to 8 bit/s for ADPCM-based calls, again depending on the background noise level. Also, the semi-lossless variant of this technique for increased steganographic bandwidth was considered in 2010 [Aoki 2010].

The PLC algorithm intended mainly for wireless VoIP systems, which rely on the specific side information that is communicated via a covert channel within the speech, was presented by Geiser et al. [2008]. To enable hidden transmission the ACELP (Algebraic Code-Excited Linear Prediction) codebook (or fixed codebook, FCB) is partitioned into sub-codebooks that uniquely identify the selected secret message's bits. Prototype AMR (Adaptive Multi-Rate)-based implementation was also described and a 2 kbit/s steganographic bandwidth was achieved experimentally.

An interesting algorithm for QIM (Quantization Index Modulation) that can be utilised for low bit-rate speech streams was introduced by Xiao et al., [2008] and later named CNV-QIM (Complementary Neighbour Vertices-Quantisation Index Modulation) [Li et al. 2012]. It is based on dividing the codebook into two parts, each representing '0' and '1', respectively. Moreover, the relationship between codewords is considered using the CNV (Complementary Neighbour Vertices) algorithm. This guarantees that every codeword is in the opposite part to its nearest neighbour, thus giving a bound of distortion. Experiments for iLBC (Internet Low Bit Rate Codec) and G.723.1 speech codecs proved that the proposed method is effective, as it only slightly decreases the speech quality whilst providing a steganographic bandwidth of 100 bit/s.

Deng et al., [2008] proposed a concept of Covert Speech Telephone (CST) that is intended to provide secure covert voice communication. This approach utilises a robust DFT (Discrete Fourier Transform) watermarking scheme to hide covert speech in the G.711-based VoIP stream. The main innovation is to use speech recognition to efficiently reduce the size of the secret information, encrypt it and hide it in the existing overt VoIP call.

An interesting study is described in [Nishimura 2009], where author proposed hiding information in the AMR-coded stream by using an extended quantisation-based method of pitch delay (one of the AMR codec parameters). This additional data transmission channel was used to extend the audio bandwidth from narrow-band (0.3–3.4 kHz) to wide-band (0.3–7.5 kHz).

Nutzinger and Wurzer [2011] introduced novel approach to speech phase coding. Typically, the original phase values are replaced with some random data. In proposed scheme the original phase values are preserved to retain high voice quality. The algorithm embeds secret data by introducing a configurable phase difference between the mean of selected chunks of the phase spectrum. Experimental results prove that the proposed method can achieve up to 12.5 bit/s while introducing almost no degradation on the voice signal, good robustness and security.

A high-capacity steganography technique based on the utilisation of inactive frames of G.723.1 speech codec was introduced by Huang et al. [2011b]. The authors prove that the inactive frames of VoIP streams are more suitable for data embedding than the active ones, thus more hidden data can be embedded in them with the same imperceptibility. They then proposed a steganographic algorithm in different speech parameters of the inactive frames for G.723.1 codec with 6.3 kbit/s bitrate. Experimental results show the solution is imperceptible and a high steganographic bandwidth up to 101 bits/frame is achieved.

The following table summarises most important characteristics of the steganographic methods outlined in subsection 3.1.

Table I. Summary of the VoIP steganography methods applied to voice payload

| | New concept for VoIP environment | New application for VoIP | Extension of existing VoIP method | Improvement of: | | |
|---|---|---|---|---|---|---|
| | | | | Undetectability | Robustness | Steganographic bandwidth |
| Wang and Wu [2007] | + | − | + | N/A | N/A | N/A |
| Dittmann et al. [2005], Aoki [2004, 2007] | − | − | + | N/A | N/A | N/A |
| Aoki [2008], Deng et al. [2008], Xiao et al. [2008], Ma et al., [2007], Xu and Yang [2009]Huang et al. [2011b] | + | − | − | N/A | N/A | N/A |
| Aoki [2003], Geiser et al. [2008], Nishimura [2009] | + | + | − | N/A | N/A | N/A |
| Takahashi and Lee [2007] | + | + | + | − | − | − |
| Tian et al. [2008], Tian et al., [2009a, 2009b], Miao and Huang [2011], Tian et al. [2011a], Tian et al. [2011b] | − | − | + | + | − | − |
| Tian et al. [2010] | − | − | + | + | − | + |
| Liu et al. [2008] | + | − | + | + | − | + |
| Wu and Yang [2006], Huang et al. [2008], Xu et al. [2011], Liu et al. [2012] | − | − | + | + | − | + |
| Aoki [2008,2010] | − | − | + | − | − | + |
| Wu et al. [2009] | − | − | + | + | + | − |
| Nutzinger and Wurzer [2011] | + | − | + | + | + | − |

**3.2 Steganographic methods applied to VoIP-specific protocols**

3.2.1. Methods that modify PDU's time relations (S2)

Utilisation of the VoIP-specific protocols as a steganogram carrier was first presented by Wang et al. [2005] (later also described in [Chen et al. 2006]). The authors proposed embedding of a 24-bit watermark into the encrypted stream (e.g., Skype call) to track its propagation through the network, thus providing its de-anonymisation. The watermark is inserted by modifying the inter-packet delay for selected packets in the VoIP stream. Authors demonstrated that depending on the watermark parameters chosen, they are able to achieve a 99% true positive and 0% false positive rate while maintaining good robustness and undetectability. However, they achieved steganographic bandwidth of only about 0.3 bit/s which is enough for the described application but rather low to perform clandestine communication.

Shah et al. [2006] inspected the use of injected jitter into VoIP packets to create a covert channel. It is intended to exfiltrate users' keyboard activity e.g., authentication credentials and the authors also prove that such an attack is feasible even when the VoIP stream is encrypted.

A new kind of information hiding technique called interference channel was introduced by Shah and Blaze [2009], which creates external interference on a shared communications medium (e.g., wireless network) in order to send hidden data. They describe an implementation of a wireless interference channel for 802.11 networks that is able to successfully transfer a steganogram over data streams (with a rather low steganographic bandwidth of 1 bit per 2.5 seconds of the call) and that is proven to be especially well suited for VoIP streams.

3.2.2. Methods that modify protocol PDU – protocol specific fields (S1)

Mazurczyk and Kotulski [2006a] proposed the use of steganography in unused fields in the RTP protocol headers and digital watermarking to embed additional information into RTP traffic to provide origin authentication and content integrity. The necessary information was embedded into unused fields in the IP, UDP and RTP protocol headers, and also into the transmitted voice. The authors later further enhanced their scheme [2006b] by also incorporating a RTCP (Real-time Transport Control Protocol) functionality without the need to use a separate protocol, thus saving the bandwidth utilised by VoIP connection.

A broader view on network steganography methods that can be applied to VoIP, to its signalling protocol, SIP with SDP (Session Description Protocol) [2008a], and to its RTP streams (also with RTCP) was presented by Mazurczyk and Szczypiorski [2008b]. They discovered that a combination of information hiding solutions provides a capacity to covertly transfer about 2000 bits during the signalling phase of a connection and about 2.5 kbit/s during the conversation phase. In 2010 Lloyd [2010] extended [Mazurczyk and Szczypiorski 2008a] by introducing further steganographic methods for SIP and SDP protocols and by performing real-life experiments to verify whether they are feasible.

Bai et al., [2008] proposed a covert channel based on the jitter field of the RTCP header. This is performed in two stages: firstly, statistics of the value of the jitter field in the current network are calculated. Then, the secret message is modulated into the jitter field according to the previously calculated parameters. Utilisation of such modulation guarantees that the characteristics of the covert channel are similar to those of the overt channel.

Forbes [2009] proposed an RTP-based steganographic method that modifies the timestamp value of the RTP header to send steganograms. The method's theoretical maximum steganographic bandwidth is 350 bit/s.

Real-life experiments with VoIP steganography on an SBC (Session Border Controller) that was acting as a gatekeeper at the borders of trust, were performed by Wieser and Röning [2010]. They were trying to establish whether SBC has some countermeasures against information hiding techniques based on SIP and RTP protocols. The results showed that it was possible to achieve a high steganographic bandwidth of even up to 569 kB/s.

Huang et al. [2011a] described how to provide efficient cryptographic key distribution in a VoIP environment for covert communication. Their proposed steganographic method is based on utilisation of the NTP field of the RTCP's SR (Sender Report) as a hidden data carrier and offered steganographic bandwidth of 54 bit/s with good undetectability.

The TranSteg (Transcoding Steganography) method that relies on the compression of the overt data in a payload field of RTP packets, in order to make free space for a steganogram, was introduced by Mazurczyk et al. [2011]. In TranSteg for a chosen voice stream, a codec that will result in a similar voice quality, but for a smaller voice payload size than the original, is found. Then, the voice stream is transcoded. At this stage, the original voice payload size is intentionally unaltered and the change of the codec is not indicated. Instead, after placing the transcoded voice payload, the remaining free space is filled with hidden data. The resulting steganographic bandwidth that was obtained using proof-of-concept implementation was 32 kbit/s while introducing delays lower than 1 ms, and still retaining good voice quality. The work was further extended by analysing the influence of speech codecs selection on TranSteg efficiency [Janicki et al. 2012a]. One of the interesting finding was that if the pair G.711/G.711.0 codecs are utilized TranSteg introduces no steganographic cost and it offers a remarkably high steganographic bandwidth, on average about 31 kbps.

In [2012] Tian et al. experimentally evaluated steganographic bandwidth and undetectability of two VoIP steganography methods proposed earlier by Huang et. al [2011a] (LSBs of NTP Timestamp field of RTCP protocol) and by Forbes [2009] (LSBs of Timestamp field of RTP protocol). Authors utilized Windows Live Messenger voice conversations system and proved that by using the first approach steganographic bandwidth of 335 bit/s can obtained and by using the second 5.1 bit/s. The latter method is also harder to detect.

3.2.3. Hybrid methods (S3)

In [Mazurczyk and Szczypiorski 2008b] a novel method called LACK (Lost Audio Packets Steganography) was introduced; it was later described and analytically analysed in [Mazurczyk and Lubacz 2010]. LACK relies on the modification of both the content of the RTP packets and their time dependencies. This method takes advantage of the fact that in RTP, excessively delayed packets are not used for the reconstruction of the transmitted data by the receiver; that is, the packets are considered useless and therefore discarded. Thus, hidden communication is possible by introducing intentional delays to selected RTP packets and by substituting the original payload with a steganogram. Practical evaluation based on the LACK prototype was presented also by Mazurczyk [Mazurczyk 2012] where the method's impact on the quality of voice transmission was investigated. The concept of LACK was further extended by Hamdaqa and Tahvildari [2011] by providing a reliability and fault tolerance mechanism based on a modified (k, n) threshold based on Lagrange Interpolation and results demonstrated that the complexity of steganalysis is increased. The "cost" for the extra reliability is a loss of some fraction of the steganographic bandwidth.

Arackaparambil et al. [2012] described a simple VoIP steganography method in which chosen RTP packets' payloads are replaced with a steganogram and the RTP header's sequence number, and/or timestamp fields are intentionally changed to make them appear as if they were excessively delayed by the network. This solution can be treated as a variation of the LACK method described above.

The following table summarises most important characteristics of the steganographic methods outlined in subsection 3.2.

Table II. Summary of the VoIP steganography methods applied to VoIP-specific protocols

|  | New concept for VoIP environment | New application for VoIP | Extension of existing VoIP method | Improvement of: | | |
|---|---|---|---|---|---|---|
|  |  |  |  | Undetectability | Robustness | Steganographic bandwidth |
| Wang et al. 2005], Shah et al. [2006], Mazurczyk and Kotulski [2006a, b], Shah and Blaze [2009], Huang et al. [2011a] | + | + | − | N/A | N/A | N/A |
| Chen et al. [2006], Lloyd [2010], Arackaparambil et al. [2012] | − | − | + | N/A | N/A | N/A |
| Mazurczyk and Szczypiorski [2008a, b], Bai et al., [2008], Mazurczyk et al. [2011] | + | − | − | N/A | N/A | N/A |
| Forbes [2009], Wieser and Röning [2010], Mazurczyk and Lubacz [2010] | − | − | + | − | − | + |
| Janicki et al. [2012a] | − | − | + | + | + | + |
| Hamdaqa and Tahvildari [2011], Tian et al. [2012] | − | − | + | + | − | − |

## 4. VOIP STEGANALYSIS

For network steganography, as well as for digital media steganography, there is still no universal, "one size fits all" detection solution, thus, steganalysis methods must be adjusted precisely to the specific information hiding technique. That is why it is hard to compare existing VoIP steganalysis methods. However, it can be noted that as a classifier typically SVM (Support Vector Machine) is utilized which is state of the art solution for steganalysis in general.

Let us consider again the hidden communication scenarios from Fig. 4. For network steganography, there are three possible localisations for a warden (W1-W3), as mentioned in Section 2. A node that performs steganalysis can be placed near the sender, or receiver of the overt communication, or at some intermediate node. Moreover, the warden can monitor network traffic in single (centralised warden), or multiple locations (distributed warden). In general, the localisation and number of locations in which the warden is able to inspect traffic greatly influences the effectiveness of the detection method.

For example, in scenario (1) from Fig. 4 – if a warden operates in an overt/covert sender's LAN (Local Area Network) then some steganographic methods (e.g., those that utilise the packets' delay) can be trivial to detect. This is because the anomaly introduced into the packets' stream will be easily spotted near the transmitter. However, if a warden is present only in the sender's or receiver's LAN, then for scenario (4) hidden communication will remain undiscovered. Similar conclusions can be drawn for a reverse case and for other communication scenarios.

Moreover, if a distributed warden is used that has access to the same traffic flow in several network localisations, then the warden's effectiveness is likely to increase. Depending on the communication scenario, steganographic modification to the network traffic can be spotted by simple comparison in two distinct locations. For example, if scenario (4) is utilised and a steganographic method like LSB is used, then by comparing the packets' payload in two different locations (e.g., in sender's LAN and on some intermediate node in the external network) it is possible to uncover steganographic traffic modification. However, it must be emphasized that, in practice, realisation of a distributed warden is hard to achieve, especially for VoIP.

Now, let us consider again the VoIP steganography classification introduced in Section 1 (Fig. 2). To verify the feasibility of methods from groups S2 and S3, experiments on real-life VoIP connections were conducted by Mazurczyk et al. [2010]. The results obtained show that from group S2 only the method that introduces intentional losses is practically applicable, but it offers low steganographic bandwidth (< 1 bit/s). The other methods that employ reordering, or modification of inter-packet delays are impractical and easy to detect. Reordering of the RTP packets was never witnessed during the experiments and inter-packet delays varied so much because of the network conditions, that applying such steganographic techniques would be a difficult task and would result (if they worked at all) in a very low steganographic bandwidth. For methods from group S3, it was concluded based on the LACK example, that it can offer potentially high steganographic bandwidth when it tries to mimic delay spikes, the characteristic formation of packets, which can lead to packet drops at the receiving end. This in turn is possible by intentionally causing such RTP packet sequences that will surely lead to jitter buffer (receiving buffer) losses by causing late packet drops, or jitter buffer overflows. Thus, an effective steganalysis method for LACK is still desirable.

As mentioned in the Section 1, the main focus of the research community in the last decade has been dedicated to developing steganographic techniques from the S1 group (see Fig. 2). This in turn resulted in the increased number of steganalysis methods developed for this group. It must be emphasised that many so called audio steganalysis methods were also developed for detection of hidden data in audio files (so called audio steganography). However, in this paper we consider only these detection methods that have been evaluated and proved feasible for VoIP.

Statistical steganalysis for LSB-based VoIP steganography was proposed by Dittmann et al. [2005]. They proved that it was possible to be able to detect hidden communication with almost a 99% success rate under the assumption that there are no packet losses and the steganogram is unencrypted/uncompressed.

Takahasi and Lee [2007] described a detection method based on calculating the distances between each audio signal and its de-noised residual by using different audio quality metrics. Then a SVM classifier is utilised for detection of the existence of hidden data. This scheme was tested on LSB, DSSS, FHSSS and Echo hiding methods and the results obtained show that for the first three algorithms the detection rate was about 94% and for the last it was about 73%.

A Mel-Cepstrum based detection, known from speaker and speech recognition, was introduced by Krätzer and Dittmann [2007] for the purpose of VoIP steganalysis. Under the assumption that a steganographic message is not permanently embedded from the start to the end of the conversation, the authors demonstrated that detection of an LSB-based steganography is efficient with a success rate of 100%. This work was further extended by [Krätzer and Dittmann 2008a] employing an SVM classifier. In [Krätzer and Dittmann 2008b] it was shown for an example of VoIP steganalysis that channel character specific detection performs better than when channel characteristic features are not considered.

Steganalysis of LSB steganography based on a sliding window mechanism and an improved variant of the previously known Regular Singular (RS) algorithm was proposed by Huang et al. [2011c]. Their approach provides a 64% decrease in the detection time over the classic RS, which makes it suitable for VoIP. Moreover, experimental results prove that this solution is able to detect up to five simultaneous VoIP covert channels with a 100% success rate.

Huang et al. [2011d] also introduced the steganalysis method for compressed VoIP speech that is based on second statistics. In order to estimate the length of the hidden message, the authors proposed to embed hidden data into a sampled speech at a fixed embedding rate, followed by embedding other information at a different level of data embedding. Experimental results showed that this solution not only allows the detection of hidden data embedded in a compressed VoIP call, but also to accurately estimate its size.

Steganalysis that relies on the classification of RTP packets (as steganographic or non steganographic ones) and utilises specialised random projection matrices that take advantage of prior knowledge about the normal traffic structure was proposed by Garateguy et al. [2011]. Their approach is based on the assumption that normal traffic packets belong to a subspace of a smaller dimension (first method), or that they can be included in a convex set (second method). Experimental results showed that the subspace-based model proved to be very simple and yielded very good performance, while the convex set-based one was more powerful, but more time-consuming.

Arackaparambil et al. [2012] analysed how in distribution-based steganalysis, the length of the window of the detection threshold and in which the distribution is measured, should be depicted to provide the greatest chance for success. The results obtained showed how these two parameters should be set for achieving a high rate of detection, whilst maintaining a low rate of false positives. This approach was evaluated based on real-life VoIP traces and a prototype implementation of a simple steganographic method and was proved efficient.

A method for detecting CNV-QIM (Complementary Neighbour Vertices-Quantisation Index Modulation) steganography in G.723.1 voice streams was described Li and Huang [2012]. This approach is to build the two models: distribution histogram and state transition model to quantify the codeword distribution characteristics. Based on these two models, feature vectors for training the classifiers for steganalysis are obtained. The technique is implemented by constructing an SVM classifier and the results show that it can achieve an average detecting success rate of 96% when the duration of the G.723.1 compressed speech bit stream is less than 5 seconds.

A steganalysis method for TranSteg based on MFCC (Mel-Frequency Cepstral Coefficients) parameters and GMMs (Gaussian Mixture Models) was developed and tested for various overt/covert codec pairs in a single warden scenario with double transcoding [Janicki et al. 2012b]. The proposed method allowed for efficient detection of some codec pairs e.g., G.711/G.726, with an average detection probability of 94.6%, Speex7/G.729 with 89.6%, or Speex7/iLBC, with 86.3% detectability. Other codecs pairs remained more resistant to detection e.g., for the pair iLBC/AMR average detection probability of 67% was achieved. Successful detection of TranSteg using the proposed steganalysis method requires at least 2 s of speech data to analyse.

## 5. CONCLUSIONS

We are currently facing an almost ubiquitous proliferation of VoIP. Skype, VoIP-ready Android phones and numerous VoIP ISPs and open source communities have resulted in a large number of users and a huge volume of traffic generated. This in turn, together with the number of protocols that form IP telephony, make it a perfect steganographic carrier that can be exploited for good (e.g., to enrich the service) or bad (e.g., to enable confidential data exfiltration) intentions. Thus, it is not surprising that clandestine communication based on steganophony is becoming reality and soon can be widely utilized with good or bad intentions.

The main focus of the research community during the last decade has been mainly dedicated to those VoIP steganography methods that modify the speech that is exchanged between the calling parties, or the codec that the speech is encoded with. Surprisingly, a lot of research effort is still devoted for improving methods like LSB. Researchers have also targeted various VoIP-specific protocols like SIP with SDP, RTP and RTCP as steganographic carriers. Little effort has been devoted to the deployment of the methods that modify the time relations between packets in the RTP stream, which is an important branch of the network steganography field. This is mainly because such solutions are rather hard to deploy practically as they usually offer low steganographic bandwidth and require synchronisation. However for certain applications like flow watermarking that do not require high steganographic bandwidth they were proved efficient and feasible [Wang et al. 2005]. An interesting research direction could be development of hybrid methods that modify both VoIP PDU's and their time relations as it has been proved that such methods offer decent steganographic bandwidth while still being undetectable [Mazurczyk and Szczypiorski 2008b]. Moreover, general techniques that improve undetectability are currently closely analysed [Tian et al. 2011b].

Of course this survey does not exhaust all of the possibilities of hidden communication that can be used with VoIP. More work will surely continue to appear on information hiding in the carried speech as well as in the VoIP-specific protocols. Obviously some of existing steganographic methods for other carriers could be successfully "transferred" to VoIP environment e.g. [Yu et al. 2007, Ling et al. 2011]. However, each method must be carefully evaluated due to specific requirements for IP telephony as a real-time service.

At the same time the steganalysis methods are present, but they are not universal and efficient enough to be practically deployed in the telecommunication network environment to perform real-time detection. Similarly like for steganographic methods some existing steganalysis solutions could be utilized also for VoIP steganography e.g. [Gianvecchio and Wang 2011; Jia et al. 2009] but as in the previous case their efficiency were not proved for real-time services.

Both, steganographic methods with the numerous ways to make them even more hidden are being developed in parallel with the methods of detection. For every VoIP steganography method developed a new steganalysis solution is sooner or later proposed. Then, even more stealthy approaches of previously known methods are developed and their detection must once again catch up. This form of "arms race" will surely not come to an end in the foreseeable future.

**ACKNOWLEDGMENTS**

Author would like to thank Prof. Krzysztof Szczypiorski from Warsaw University of Technology (Poland) for helpful comments and remarks.

# REFERENCES

AOKI, N. 2003. A Packet Loss Concealment Technique for VoIP using Steganography, In Proc. of International Symposium on Intelligent Signal Processing and Communication Systems (ISPACS 2003), Awaji Island, Japan, 470-473

AOKI, N. 2004. VoIP packet loss concealment based on two-side pitch waveform replication technique using steganography, In Proc. of TENCON 2004, Thailand, Vol. 3, 52 – 55

AOKI, N. 2007. Potential of Value-Added Speech Communications by Using Steganography, In Proc. of Intelligent Information Hiding and Multimedia Signal Processing (IIHMSP 2007), 251 – 254

AOKI, N. 2008. A technique of lossless steganography for G.711 telephony speech. In Proc. of Int Conf Intelligent Information Hiding and Multimedia Signal Processing (IIHMSP 2008), Harbin, China, 608–611

AOKI, N. 2009. Lossless Steganography Techniques for IP Telephony Speech Taking Account of the Redundancy of Folded Binary Code, In Proc. of 5th International Joint Conference on INC, IMS and IDC, 1689 - 1692

AOKI, N. 2010. A Semi-Lossless Steganography Technique for G.711 Telephony Speech, In Proc. of Sixth International Conference on Intelligent Information Hiding and Multimedia Signal Processing (IIHMSP 2010), Darmstadt, Germany, 534 - 537

ARACKAPARAMBIL, C., YAN, G., BRATUS, S., CAGLAYAN, A. 2012. On Tuning the Knobs of Distribution-based Methods for Detecting VoIP Covert Channels, In Proc. of Hawaii International Conference on System Sciences (HICSS-45), Hawaii, 2431-2440

BAI, L. Y., HUANG, Y., HOU, G., XIAO, B. 2008. Covert channels based on jitter field of the RTCP Header. In Proc. of Int Conf Intelligent Information Hiding and Multimedia Signal Processing (IIHMSP 2008), Harbin, China, 1388 - 1391

BENDER, W., GRUHL, D., MORIMOTO, N., LU, A. 1996 Techniques for data hiding. IBM. Syst J 35(3/4):313–336

CHEN, S., WANG, X., JAJODIA, S. 2006. On the Anonymity and Traceability of Peer-to-Peer VoIP Calls. IEEE Network, 20(5):32–37

DENG, Z., SHAO, X., YANG, Z., ZHENG, B. 2008. A novel covert speech communication system and its implementation, Journal of Electronics, Vol. 25 No. 6, 737-745

DITTMANN, J., HESSE, D., HILLERT, R. 2005. Steganography and steganalysis in voice-over IP scenarios: operational aspects and first experiences with a new steganalysis tool set. In: Proc of SPIE 2005, Vol. 5681, Security, Steganography, and Watermarking of Multimedia Contents VII, San Jose, USA, 607–618

DoD „Orange Book", National Computer Security Center, Trusted Computer System Evaluation Criteria, Tech. Rep. DOD 5200.28-STD, National Computer Security Center, Dec. 1985

DRUID. 2007. Real-time steganography with RTP. Technical Report http://www.uninformed.org/?v=8&a=3&t=pdf

FORBES, C.R. 2009. A New covert channel over RTP. MSc thesis, Rochester Institute of Technology. https://ritdml.rit.edu/bitstream/handle/1850/12883/CForbesThesis8-21-2009.pdf?sequence=1

FRIDRICH, J., (2010) Steganography in digital media – Principles, Algorithms, and Applications, Cambridge University Press, ISBN 978-0-521-19019-0

GARATEGUY, G., ARCE, G., PELAEZ, J. 2011. Covert Channel detection in VoIP streams, In Proc. of 45th Annual Conference on Information Sciences and Systems (CISS), 1-6

GEISER B., MERTZ F., VARY, P. 2008. Steganographic Packet Loss Concealment for Wireless VoIP, In Proc. of Conference on Voice Communication (SprachKommunikation), 1-4

GIANVECCHIO, S., WANG, H. 2011. An Entropy-Based Approach to Detecting Covert Timing Channels, In IEEE Transactions on Dependable and Secure Computing, Vol. 8, No. 6

HAMDAQA M., TAHVILDARI L. 2011. ReLACK: A Reliable VoIP Steganography Approach, In Proc. of Fifth International Conference on Secure Software Integration and Reliability Improvement (SSIRI 2011), Korea, 189 - 197

HUANG Y., XIAO B., XIAO H. 2008. Implementation of Covert Communication based on Steganography, International Conference on Intelligent Information Hiding and Multimedia Signal Processing (IIH-MSP 2008), Harbin, China, 1512 - 1515

HUANG, Y., YUAN, J., CHEN, M., XIAO, B. 2011a. Key Distribution over the Covert Communication Based on VoIP, Chinese Journal of Electronics, Vol. 20, No. 2, 357-360

HUANG Y., TANG, S., YUAN, J. 2011b. Steganography in Inactive Frames of VoIP Streams Encoded by Source Codec, IEEE Transactions on information forensics and security, Vol. 6, No. 2, 296-306

HUANG, Y., TANG, S., ZHANG, Y. 2011c. Detection of covert voice-over Internet protocol communications using sliding window-based steganalysis, IET Communications, Vol. 5, Iss. 7, 929–936

HUANG, Y., TANG, S., BAO, C., YIP, YAU J. 2011d. Steganalysis of compressed speech to detect covert voice over Internet protocol channels. IET Information Security, 5 (1), 26-32

JANICKI, A., MAZURCZYK, SZCZYPIORSKI, K.. 2012a. Influence of Speech Codecs Selection on Transcoding Steganography, In: Computing Research Repository (CoRR), abs/1201.6218, arXiv.org E-print Archive, Cornell University, Ithaca, NY (USA), URL: http://arxiv.org/abs/1201.6218

JANICKI, A., MAZURCZYK, SZCZYPIORSKI, K.. 2012b. Steganalysis of Transcoding Steganography, In: Computing Research Repository (CoRR), abs/1210.5888, arXiv.org E-print Archive, Cornell University, Ithaca, NY (USA), URL: http://arxiv.org/abs/1210.5888

JIA, W., TSO F. P., LING, Z., FU, X., XUAN, D., YU, W. 2009. Blind Detection of Spread Spectrum Flow Watermarks, In Proc. of IEEE INFOCOM 2009, 2195 – 2203

KRÄTZER, C., DITTMANN, J., VOGEL, T., HILLERT, R. 2006. Design and evaluation of steganography for Voice-over-IP, In Proc. of IEEE Int Symp Circuits and Systems (ISCAS 2006), Kos, Greece

KRÄTZER, C., DITTMANN, J. 2007. Mel-Cepstrum Based Steganalysis for VoIP-Steganography, In Proc. of the 19th Annual Symposium of the Electronic Imaging Science and Technology, SPIE and IS&T, San Jose, California, USA

KRÄTZER, C., DITTMANN, J. 2008a. Pros and Cons of Mel-cepstrum based Audio Steganalysis using SVM Classification, In Lecture Notes on Computer Science, vol. LNCS 4567, 359–377

KRÄTZER, C., DITTMANN, J. 2008b. Cover Signal Specific Steganalysis: the Impact of Training on the Example of two Selected Audio Steganalysis Approaches, In Proc. of SPIE-IS&T Electronic Imaging, SPIE Vol. 6819

KUNDUR, D., AHSAN, K. 2003. Practical Internet Steganography: Data Hiding in IP, In Proc. of the Texas Workshop on Security of Information Systems, USA

LAMPSON, B.W. 1973. A note on the confinement problem, Comm. ACM 16, 10, 613-615

LI, S., HUANG, Y. 2012. Detection of QIM Steganography in G.723.1 Bit Stream Based on Quantization Index Sequence Analysis, Journal of Zhejiang University Science C (Computers & Electronics), Vol. 13, Iss. 8, 624-634

LING, Z., FU, X., JIA, W., YU, W., XUAN, D. 2011. A Novel Packet Size Based Covert Channel Attack against Anonymizer, In the mini-conference conjunction with IEEE International Conference on Computer Communications (INFOCOM), 186-190

LIU, L., LI, M., LI, Q., LIANG, Y. 2008. Perceptually Transparent Information Hiding in G.729 Bitstream, In Proc. of the 4th International Conference on Intelligent Information Hiding and Multimedia Signal Processing, Harbin, China, 406-409.

LIU, J., ZHOU, K., TIAN, H. 2012. Least-Significant-Digit Steganography in Low Bit-Rate Speech. In proc. of the 47th IEEE International Conference on Communications (ICC), Ottawa, Canada, 1-5

LLOYD, P. 2010. An Exploration of Covert Channels Within Voice Over IP, MSc thesis, Rochester Institute of Technology, https://ritdml.rit.edu/bitstream/handle/1850/12241/PLloydThesis5-4-2010.pdf?sequence=1

LU, C. S. 2005. Multimedia security: steganography and digital watermarking techniques for protection of intellectual property, Idea Group Publishing, ISBN 1-59140-192-5

LUBACZ, J., MAZURCZYK, W., SZCZYPIORSKI, K. 2008. Hiding Data in VoIP, In Proc of: The 26th Army Science Conference (ASC 2008), Orlando, Florida, USA

LUBACZ, J., MAZURCZYK, W., SZCZYPIORSKI, K. 2010. Vice over IP, IEEE Spectrum, 40–45

LUCENA, N. B., PEASE, J., YADOLLAHPOUR, P, CHAPIN, S. 2004. Syntax and Symantics-Preserving Application-Layer Protocol Steganography, In Proc. of IH 2004, LNCS 3200, pp. 164-179

MA, L., WU, Z., YANG, W. 2007. Approach to Hide Secret Speech Information in G.721 Scheme, In Proc. of ICIC 2007, LNCS 4681, 1315–1324

MAZURCZYK W., Z. KOTULSKI Z. 2006a. New VoIP traffic security scheme with digital watermarking, In Proceedings of The 25-th International Conference on Computer Safety, Reliability and Security SafeComp 2006, Lecture Notes in Computer Science 4166, pp. 170 – 181

MAZURCZYK, W., KOTULSKI, Z. 2006b. New security and control protocol for VoIP based on steganography and digital watermarking, In: Proc 5th Int Conf Computer Science – Research and Applications (IBIZA 2006), Kazimierz Dolny, Poland

MAZURCZYK, W., SZCZYPIORSKI, K. 2008a. Covert channels in SIP for VoIP signalling. In Proc. of 4th Int Conf Global E-security, London, United Kingdom, 65–70

MAZURCZYK, W., SZCZYPIORSKI, K. 2008b. Steganography of VoIP Streams. In: Meersman R, Tari Z (eds) OTM 2008, Part II – Lecture Notes in Computer Science (LNCS) 5332, Springer-Verlag Berlin Heidelberg, Proc On The Move Federated Conferences and Workshops: 3rd Int Symp Information Security (IS'08), Monterrey, Mexico, 1001–1018

MAZURCZYK, W., LUBACZ, J. 2010. LACK – a VoIP steganographic method, Telecommun Syst: Model Anal Des Manag 45(2/3), 153-163

MAZURCZYK, W., CABAJ, K., SZCZYPIORSKI, K. 2010. What are suspicious VoIP delays?, In: Multimedia Tools and Applications, Vol. 57, No 1, 109-126

MAZURCZYK, W., SZAGA, P., SZCZYPIORSKI, K. 2011. Using transcoding for hidden communication in IP telephony, In: Computing Research Repository (CoRR), abs/1111.1250, arXiv.org E-print Archive, Cornell University, Ithaca, NY (USA)
http://arxiv.org/abs/1111.1250

MAZURCZYK, W. 2012. Lost Audio Packets Steganography: A First Practical Evaluation, International Journal of Security and Communication Networks, John Wiley & Sons, ISSN: 1939-0114, DOI: 10.1002/sec.502 (to appear)

MIAO, R., HUANG, Y. 2011. An approach of covert communication based on the adaptive steganography scheme on Voice over IP, IEEE Int Conf Communications (ICC 2011), 1-5

MURDOCH, S., LEWIS, S. 2005. Embedding covert channels into TCP/IP, In Proc. of Inf. Hiding, 247–266

NISHIMURA, A. 2009. Steganographic bandwidth extension for the AMR codec of low-bit-rate modes. In Proc. of Interspeech 2009, Brighton, UK, 2611–2614

NUTZINGER M., FABIAN, C. MARSCHALEK, M. 2010. Secure Hybrid Spread Spectrum System for Stegnanography in Auditive Media, In Proc. of IEEE International Conference on Intelligent Information Hiding and Multimedia Signal Processing 2010, Darmstadt, Germany, 78-81

NUTZINGER M., WURZER, J. 2011. A Novel Phase Coding Technique for Steganography in Auditive Media, In Proc. of IEEE International Conference on Availability, Reliability and Security, Vienna, Austria, 91-98.

PETITCOLAS F, ANDERSON R, KUHN M 1999. Information hiding—a survey IEEE. Special Issue on Protection of Multimedia Content, Vol. 87, Iss. 7, 1062 - 1078

ROSENBERG, J., SCHULZRINNE, H., CAMARILLO, G., JOHNSTON, A. 2002. SIP: Session Initiation Protocol. IETF, RFC 3261

SCHULZRINNE, H., CASNER, S., FREDERICK, R., JACOBSON, V. 2003. RTP: A transport protocol for real-time applications. IETF, RFC 3550

SERVETTO, S. D., VETTERLI, M. 2001. Communication Using Phantoms: Covert Channels in the Internet, In Proc. of IEEE International. Symposium on Information Theory (ISIT)

SHAH, G., MOLINA, A., BLAZE, M. 2006. Keyboards and Covert Channels. In Proc. of 15th USENIX Security Symposium, 59–75

SHAH G., BLAZE M. 2009. Covert channels through external interference, In Proc. of the 3rd USENIX conference on Offensive technologies, Montreal, Canada

SIMMONS, G.J. 1984. The prisoner's problem and the subliminal channel. In Advances in Cryptology: Proceedings of CRYPTO '83. Plenum Press, Santa Barbara, California, USA

TAKAHASHI, T., LEE, W. 2007. An assessment of VoIP covert channel threats. In Proc. of 3rd Int Conf Security and Privacy in Communication Networks (SecureComm 2007), Nice, France, 371 - 380

TIAN, H., ZHOU, K., HUANG, Y., FENG, D., LIU, J. 2008. A Covert Communication Model Based on Least Significant Bits Steganography in Voice over IP, In Proc. of 9th International Conference for Young Computer Scientists, China, 647-652

TIAN, H., ZHOU, K., JIANG, H., LIU, J., HUANG, Y., FENG, D. 2009a. An M-sequence based steganography model for voice over IP, In Proc. of IEEE Int Conf Communications (ICC 2009), Dresden, Germany, 1–5

TIAN, H., ZHOU, K., JIANG, H., LIU, J., HUANG, Y., FENG, D. 2009b. An adaptive steganography scheme for voice over IP, IEEE Int Symp Circuits and Systems (ISCAS 2009), Taipei, Taiwan, 2922-2925

TIAN, H., ZHOU, K., FENG, D. 2010. Dynamic matrix encoding strategy for voice-over-IP steganography, J. Cent. South Univ. Technol. 17: 1285−1292

TIAN, H., JIANG, H., ZHOU, K., FENG, D. 2011a. Adaptive partial-matching steganography for voice over IP using triple M sequences, Computer Communications Journal, 34, 2236–2247

TIAN, H., JIANG, H., ZHOU, K., FENG, D. 2011b. Transparency-orientated Encoding Strategies for Voice-over-IP Steganography, The Computer Journal, The Computer Journal, 2012, 55(6): 702−716

TIAN, H., GUO, R., LU, J., CHEN, Y. 2012. Implementing Covert Communication over Voice Conversations with Windows Live Messenger, Advances in information Sciences and Service Sciences (AISS), Vol. 4, No. 4, 18-26

WANG, C., WU, W. 2007. Information hiding in real-time VoIP streams, In Proc. of 9th IEEE Int Symp Multimedia (ISM 2007), Taichung, Taiwan, 255–262

WANG, X., CHEN, S., JAJODIA, S. 2005. Tracking anonymous peer-to-peer VoIP calls on the internet, In Proc. of the 12th ACM Conference on Computer and Communications Security (CCS '05), pp. 81-91

WIESER, C., RÖNING, J. 2010. An Evaluation of VoIP covert channels in an SBC setting, In Proc. of Security in Futures – Security in Change Conference, Turku, Finland, 54-58

WU, Z. J., GAO, W., YANG, W. 2009. LPC parameters substitution for speech information hiding, The Journal of China Universities of Posts and Telecommunications, 16(6): 103–112

WU, Z. J., YANG, W., YANG, Y. X. 2003. ABS-based Speech Information Hiding Approach, IEE Electronics Letters, 39, 1617-1619

WU, Z., YANG, W. 2006. G.711-Based Adaptive Speech Information Hiding Approach, In Proc. of ICIC 2006, LNCS 4113, 1139 – 1144

XIAO, B., HUANG, Y., TANG, S. 2008. An Approach to Information Hiding in Low Bit-rate Speech Stream, In Proc. of Global Telecommunications Conference (GLOBECOM 2008), USA, 1-5

XU, T., YANG, Z. 2009. Simple and effective speech steganography in G.723.1 low-rate codes, In Proc. of Int Conf Wireless Communications & Signal Processing WCSP 2009, Nanjing, China, 1–4

XU, E., LIU, B., XU, L., WEI, Z., ZHAO, B., SU, J. 2011. Adaptive VoIP Steganography for Information Hiding within Network Audio Streams, In Proc. of Int. Conference on Network-Based Information Systems, Tirana, Albania, 612 – 617

YU, W., FU, X., GRAHAM, S., XUAN, D., ZHAO, W. 2007 DSSS-Based Flow Marking Technique for Invisible Traceback, In proc. of IEEE S&P 2007, 18-32

ZANDER, S., ARMITAGE, G., BRANCH, P. 2007. A survey of covert channels and countermeasures in computer network protocols, IEEE Commun Surv Tutor 9(3):44–57